\documentclass{PoS}

\title{Fast 3C~279 $\gamma$ flares by a merging medium size black hole jet aligned to the AGN one by tidal torque?}

\ShortTitle{
3C~279 $\gamma$-flare by merging BH -- AGN aligned jets
}

\author{Daniele Fargion\\
        Physics Department, Rome University 1 \\
        INFN Rome1, Ple. A. Moro 2, 00185, Rome, Italy\\
        MIFP, Mediterranean Institute of Fundamental Physics - Via Appia Nuova 31, 00040 Marino (Rome), Italy.\\
        E-mail: \email{daniele.fargion@roma1.infn.it}}

\author{Pietro Oliva\\
        Department of Electrical Engineering, Niccol\`o Cusano University, Via Don Carlo Gnocchi 3, 00166 Rome, Italy\\
        MIFP, Mediterranean Institute of Fundamental Physics - Via Appia Nuova 31, 00040 Marino (Rome), Italy.\\
        E-mail: \email{pietro.oliva@unicusano.it}}

\abstract{The shorter-than-Schwarzschild 3C 279 variability flare on June 2015 is very puzzling. Its nature cannot be due to any NS merging nor to a medium sized ($\sim10^2$~M$_\odot$) BH collapse. Our preliminary model is based on the long-life ($\sim10^7$ s) merging of a medium size BH ($\sim10^2$~M$_\odot$) spiralling toward the largest AGN one, that is dragging by tidal torques  the medium size BH jet along the main AGN 3C 279 one. The tidal torque is aligning both jets toward Earth. The twin overlapping blazars may offer at once a long and a short scale variability consistent with the surprising Fermi discovers.}

\FullConference{XI Multifrequency Behavior of High Energy Cosmic Sources  Workshop,\\
		25-30 May 2015\\
		Palermo, Italy}

\begin{document}
\section{Introduction}
In  very recent years we have been aware of very rapid AGN flare, in time scale comparable or even less or much less than
the same AGN Black Hole time scale (see for instance the PKS 2155-304
\cite{2007ApJ...664L..71A}). One of the most amazing, hard and fast variability occurred for
the peculiar radio galaxy IC 310: its peak brightness variability during
the exceptional flare of Nov. 2012 \cite{2014Sci...346.1080A, 2015arXiv150805031E} had time scales decay as short as 4.8  minutes, while the characteristic AGN mass BH  and time scale should be as large  as 16 minutes until an hour or two (because of the uncertain BH mass).
These events are puzzling also for their hardness (TeV peak energy) and for their unexpected inner tripled variability.
In more recent events, on 2015 June 16, the Fermi-LAT detector has been blazed by a very hard gamma rain \cite{nasa2015.10.07, 2015ApJ...808L..48P} that almost reached at  peak brightness the characteristic isotropic Gamma Ray Burst's (GRB) luminosity ($\sim10^{49}$ erg s$^{-1}$). The subject is very hot and under several investigations \cite{2016ApJ...824L..20A, 2016MNRAS.457.3535Z}. We remind that
the 3-day flashes had an orbital instrumental (terrestrial-Kepler) variability of about 95.4 minutes \cite{2016ApJ...824L..20A}.

The 3C~279 has an estimated Black Hole (BH) mass of nearly half a billion ($3\,\textup{--}\,8\cdot10^8$ M$_\odot$) solar masses  \cite{2015ApJ...807...79H}.
Incidentally, the corresponding 3C~279 Schwarzschild radius $r_{s_{\mathrm{3C~279}}}$ and time scale ($t_{\mathrm{3C~279}}\sim5\cdot10^3$ s) is comparable to the same terrestrial orbital one; nevertheless, the last spiralling orbit around the BH is nearly an order of magnitude longer.
The novel, surprising (2015) discover of a huge and fast 3C~279 blazing flare, almost a hundred times brighter than most luminous AGN flashes, led us to the following question: may the observed shortest (five minutes) 3C 279 variability  \cite{2016ApJ...824L..20A} be consistent with the huge 3C~279 mass and its extremely longer Kepler BH spiral times ($\sim3\cdot10^4$~s)?
 This query has an analogy in short scale GRBs events whose time scales of ms or less are somehow  comparable  with the expected source mass (tens M$_\odot$ or more). Based on GRB's energetics, the simple proposal we already considered since more than two decades (with a recent major update, see \cite{2016arXiv160500177F} and references therein), is assuming the origin of GRBs as a modulated thin spinning-precessing $\gamma$-jet, whose blazing are explaining the light-curves features through geometry jet alignment and whose narrow beam explain apparent fast variability; other various GRB puzzles such as the absence of correlation of IceCube highest energy neutrinos with the sources \cite{2012NIMPA.692..174F, 2012APh....35..354P} found a possible answer in our recent paper \cite{2016arXiv160500177F}. The assumed GRB's jet is persistent (while spinning and precessing) and it is fed by a companion binary accretion disk blazing  only while on-axis to us.

This precessing-jet model works as in analogy with nearer binary systems like the well known SS433 whose output is much lower (than GRB ones) and whose geometry pointing is at present times not illuminating us. This object might then appear as a Soft Gamma Repeater (SGR) if it would be aligned with the Earth. Now, the same understanding might be suggested for fast 3C~279 event. The thin jet variability thanks to its narrow opening angle (related to its large, thousands, Lorentz factor) might explain at once the mysterious behavior of such a fast variability (of about two orders of magnitude briefer than BH one) in such a shortest time scales ($\sim5$ minutes or $\sim3\cdot10^2$~s) as well as the apparent modulation in GRB sample luminosity by a factor above hundred million times. Unfortunately, this simple description is  not easy to be accepted for such huge BH. There is, indeed, a remarkable difference between a GRB from  Neutron Star (NS-BH merging), jet feeding and precessing, and the 3C~279 jet variability: the mass size in each object category and the consequent probability of the merge to happen while pointing to us. AGN jets are billion times more rare than micro GRB jet ones.

\section{Time scales and probability in precessing blazing GRB jet}

Let us be more quantitative with beaming and the consequent probability to be shined: a NS-BH or NS-NS collapse is not such a rare event in our universe. It happens possibly once every a hundred years per galaxy in case of NS-NS or below for larger NS-BH masses. Because of the huge number of far galactic sources in the cosmic volume, even under the extreme GRB beaming of $\mathrm{d}\theta\sim\gamma^{-1}\sim10^{-4}$  and  $\frac{\Delta\Omega}{\Omega}\sim10^{-8}$, the probability to be originally pointing toward us at birth (for the first tens of seconds) is just rare as 0.1 per year. However, the GRB persistence and precessing along its spread angle activity is much larger than the one-shot (Fireball event) or than the apparent observed GRB blazing time: its characteristic duration time is, in our model,  as long as $t_0\sim10^4\,\textup{--}\,10^6$ s and its decay the order of magnitude of hours-days \cite{2014arXiv1408.0227F}, while its decay by a later time scale, following a power law like $\left(t/t_0\right)^{-1}$, makes GRBs jet active at lower and lower jet energies, lasting years or thousands years, still observable at nearer cosmic distances  as  short GRB (sGRB) or even at weaker energies, but nearer distances as galactic SGRs.

The probability of a similar ``large scale'' (AGN+GRB) event is not totally impossible but it is even much more rarer because AGN are much rarer than microquasars; consider just only one giant central BH versus probable millions medium BHs and hundred millions common NS for each galaxy. Indeed, the very recent LIGO report  \cite{PhysRevLett.116.061102} of medium-size BH merging ($\gtrsim20\,\textup{--}\,60$~M$_\odot$) is favoring the possibility of a large and unexpected population of such medium size objects (see also \cite{2016arXiv160707456T}). As we foresaw \cite{2016arXiv160309639F}, these BH-BH collapses are in general only observable by their gravitational waves, but still unobservable (at least in the next decade) by their electromagnetic counterpart. This because the improbable alignment of a tiny BH jet toward us while collapsing (in a spherical GWs signal) with  a companion BH. Nevertheless, the NS-BH model described very recently \cite{2016arXiv160500177F} may suggest that similar merging NS-BH$_{\mathrm{AGN}}$ may also occur. In a first approximation this is correct, but such a NS-BH$_{\mathrm{AGN}}$ collapse (if exploding at orbit in isotropic way) would shine as a diluted thermal supernova (SN) in day and day re-brightening, not as a hard, sharp and fast GeV gamma flash. Of course, one may imagine that a NS merging is a powerful source of energy able to increase sharply the AGN jet, but the time scale of such increase has much larger duration $\gtrsim r_{s_{\mathrm{3C~279}}}/c$ than the observed few minute increase. This is true both for a NS-BH$_{\mathrm{AGN}}$ frontal impact ($t_d\gtrsim10^4$ s) as well as for a longer spiral collapse ($t_s\sim10^5$ s). In conclusion the 5 minutes 3C~279 increase has not a natural connection with merging time for NS-BH$_{\mathrm{AGN}}$ collapse. One may therefore re-consider the sudden increase of luminosity by a tiny deflection of the main AGN jet, because of a NS collapse. These possibilities exist, but the mass ratio $\frac{m_{\mathrm{NS}}}{m_{\mathrm{AGN}}}\sim10^{-8}\,\textup{--}\,10^{-9}$ suggests only a consequent  tiny $\sim10^{-8}\,\textup{--}\,10^{-9}$ radiant deflection: a too small bending for such a high ($\sim40$) luminosity increase.
See also \cite{2013ApJ...774..142B} and \cite{{Borra92}, 2010A&A...511L...6B} as an interesting example of spectroscopy technique for spectral modulations.

\subsection{A Larger BH $10^5$ M$_{\odot}$  capture by main  blazing AGN BH ?}

From the previous arguments one may suggest that the main huge AGN Jet bending might be deflected effectively by a companion heavier (than hundred solar mass) BH. The fast bending and the narrow beam may be acceptable. This is (almost) possible. A very heavy BH of nearly $\gtrsim10^5$ M$_{\odot}$ can be able to deflect, during  its collapse toward the billion solar masses AGN BH, the AGN jet by a $10^{-4}$ radiant, which is almost necessary to explain the remarkable 3C~279 increase of intensity. However the main time scales are possibly modulated by the longer and longer BH Schwarzschild time scales of the binary  system as well by the relativistic redshift dilution at the AGN BH surface. Nevertheless the probability to observe, while being illuminated by the  AGN jet, such a huge  $10^5$ M$_{\odot}$ BH capture seems quite tiny requiring at least a vast (unexpected) population of such medium-large mass BHs of $10^5$ M$_{\odot}$.
Anyway there are some old and recent proposals for a primordial population of low giant BH \cite{MIDI_BH_2012},  $10^4$-$10^6$ M$_{\odot}$  whose eventual coexistence in most galaxies may explain the presence of binary BH systems in many galactic cores and the eventual AGN jet bending  by a needed $10^{-4}$ radiant. Here for the moment we neglect this tuned possibility.

\subsection{New Astronomy by GWs, UHECR and UHE Neutrino? }

One of the most popular Astronomy to be born is the one based on Highest Energy Neutrinos. IceCube provided hints for it.
The other recent GW detection at tens-hundred Solar mass BH opened a road map to a novel astronomy, if soon or later a better angular resolution will mark the sky source.
However, we did recently dismiss such an easy GW-$\gamma$ connection \cite{2016arXiv160309639F}. The most exciting astronomy, based on un-deflected ultra-relativistic (UHECR) protons, is in discussion: even we didn't agree with a nucleon composition. We claimed a Light Nuclei UHECR \cite{2009NuPhS.190..162F} that might be disentangled by differential spectroscopy of air-showering at the horizons \cite{2006PrPNP..57..384F}.

The light nuclei might cluster within the observed smeared hot spot toward Cen-A \cite{2010PrPNP..64..363F} and other possible sources \cite{2015EPJWC..9908002F}.
Such an angular discrimination of airshower  might co-work from top mountain to disentangle upward tau neutrino airshower at PeVs-EeVs energies \cite{2004NuPhS.136..119F}.
The appealing idea might come to share AGN flares with future LISA GWs detection. Unfortunately the GW due to medium BH collapse to AGN is ruled by AGN BH time scales. The GW output is quadratic to third power quadrupole derivative leading to $\omega^{6}$ factor, that it is too small for a medium size mass and too slow even for LISA frequency windows.
The natural consequence of such a huge  3C~279 gamma hardness and luminosity has driven some authors \cite{2016arXiv160506119H} to foresee a UHE neutrino-$\gamma$ correlation, as soon as the IceCube blind data will be unfolded. At the present we do not totally  deny this possibility, in particular if the AGN (of very probably  hadronic nature) is itself the unique engine shining toward us. However, in our present model we imagine that the inner fast brightest gamma flashes are produced by a small medium size jet whose most probable nature is electronic and not hadronic one. Therefore, our prediction is the absence of UHE neutrino correlated to the actual 3C~279 giant flashes.

\section{A twin jet blazing at once?}
 \begin{figure}[t]
 \centering
 \includegraphics[scale=.4]{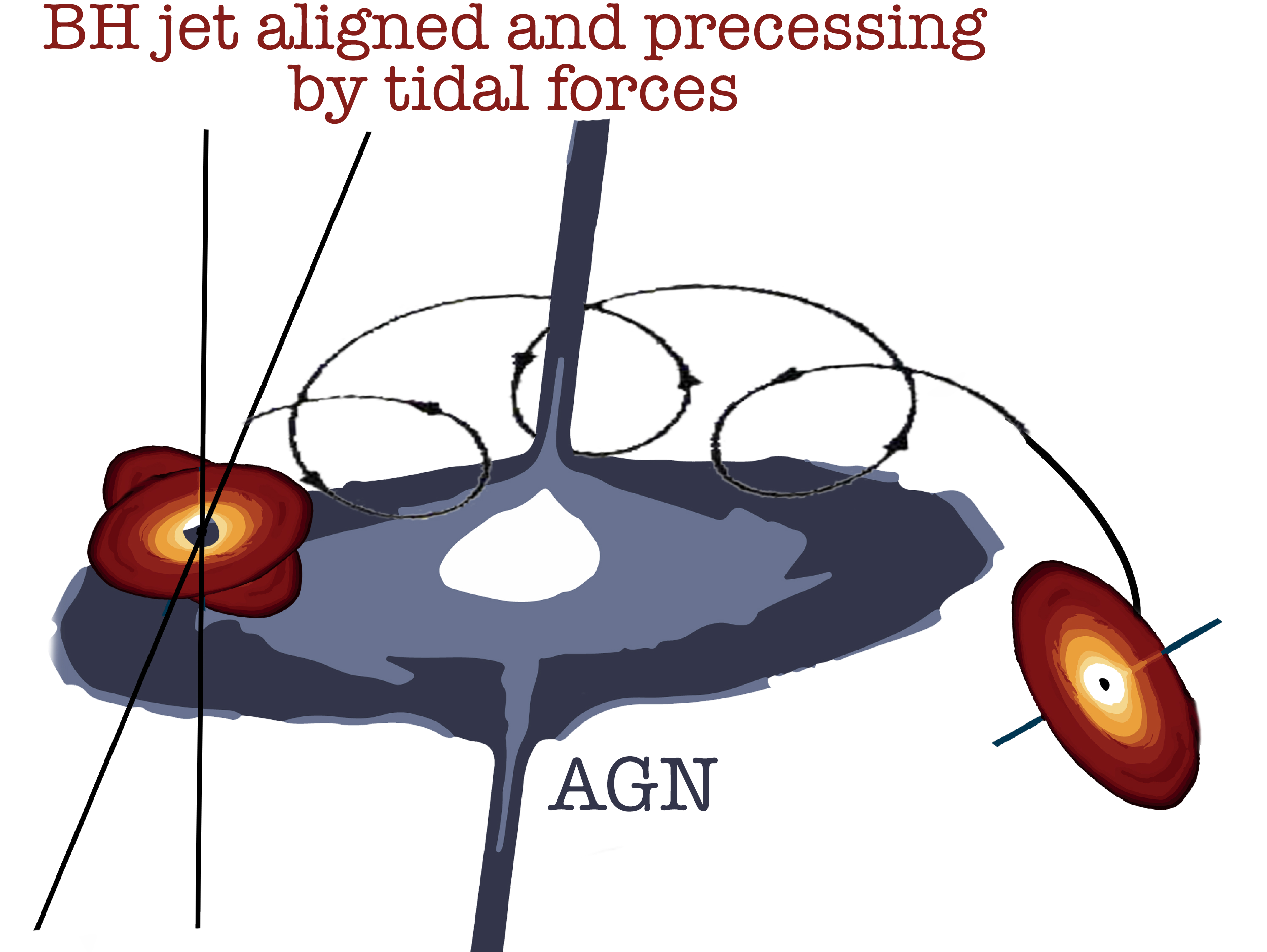}
\caption{The present picture describes the first and late stages of a spiral small scale AGN-jet, hundred solar masses, that is tidally forced by the AGN spin to align toward the Earth in final collapsing stages: the complementary blazing of the two jets might be the cause of slow and fast variability as in 3C 279  giant flare on June 2015.}
\end{figure}\label{fig:01}

The possibility that a medium-size mass ($10^2\,\textup{--}\,10^3$ M$_\odot$)  BH's jet shines, producing a long apparent GRB has been discussed recently in \cite{2016arXiv160500177F} where we tried to unify (for more at least a decade, see \cite{2007sngh.conf..133F}) long and short GRBs as well. The coincident blazing of a persistent AGN such as 3C ~279 with such a shining GRB seems extremely improbable. However, the narrow orbit between the medium BH jet and the AGN jet suffers the tidal torque spin-spin interaction \cite{2001PhRvL..87w1101P, 2016MNRAS.455.1946F, 2015arXiv151207969P}, whose minimal energy configuration is the aligned twin spin (in analogy with the electrons spin alignment in atoms).
Once again, let's clarify the sentence above: the appearance of a GRB nearby an active AGN requires the same blazing jet of a modest BH ($1\,\textup{--}\,10^2$ M$_\odot$) whose spin should be collinear with the huge AGN one. The 3C~279 is one of the rarest AGN brightening at highest level in the gamma sky, and it is probably the most collinear blazar toward us among billions AGN in the universe.
The 3C~279 long life (tens of years) blazing activity is very likely due to a slow accreting disk powered by millions stars merging along the AGN itself, feeding the jet.
 Now, the fast brief minutes blazing activity, we suggested, is due to the whole jet structure bending. Furthermore, as we mentioned above, this is quite difficult to occur because of the large $r_{s_{\mathrm{3C~279}}}/c$ time scale and the huge AGN angular momentum inertia with respect to a few hundred solar masses BH collapse. The alternative we are considering in the present article is the merging of a medium size ($\sim$ hundred M$_\odot$) BH jet whose spin pointing geometry is ``forced and frozen'' for a while to shine in the collinear direction with the parental AGN jet. This forced collinear state is possible by General Relativity (GR) effects (gravitational LT-like) and by dissipation processes during the medium BH spiralling along 3C~279. This rare phenomenon (too short variability for the AGN blazing respect to the BH Schwarzschild time scale) has some additional remarkable examples like IC~310.

We concentrated our attention here to the June 2015 flare event (the longest) because of its exceptional brightness and its sharp variability. We suggested that such main AGN jet torque the secondary medium BH thin jet  along the same terrestrial direction.  Nevertheless, it is quite possible that in general many more medium BH  are sinking into - AGN-BH
where their  jets are flashing not to us, but elsewhere. In general AGN jet and medium BH jet spin axis might be, at the initial stages, totally unrelated.
  Here we remind that the overlapping of the (long life $\sim 10^6\,\textup{--}\,10^7$ s)  rare collapse of a medium BH collinear with the AGN is favored and amplified by the gravity viscous alignment during the nearest spiralling orbits. The characteristic Kepler spiral time  for a half a billion BH is about $\sim1.8\cdot10^5$ s but the plasma viscosity and the whole process of alignment may require longer times that may reach months ($\sim10^6\,\textup{--}\,10^7$ s).
This novel understanding of AGN fast flashes results is somehow related to the new LIGO GW discovers and statistics. If the rate of medium BH collapse is not too rare (as it appears by LIGO's one-year records), the consequent AGN  BH - medium BH collapse might be quite frequent. The more unexpected twin alignment  considered in present paper has been just an additional phenomenon foresaw by GR \cite{1973ApJ...185..635T, 2001PhRvL..87w1101P}.

\section{Conclusions: Twin jets aligned by torque and drag forces}

A century after Einstein  wrote about geometrical gravity laws, after half a century of search of GWs, the recent LIGO-VIRGO discover of BH-BH merging waves made a great confirmation of all Einstein visions. However, his (and most of our) earlier fears about BH existence have been already ruled out by last decades presences of AGN flares by their luminosity, variabilities and orbital trajectories. The million-to-billion solar masses BH in AGN sky became a reality in the last half a century. The last year 3C~279 peculiar minutes flare may witness a new chapter of events: the collinear jet--jet collapse by BH AGN and medium BH mergers. Naturally, our proposal is only a first preliminary attempt to rule out the puzzling 3C~279 flashes. More details and verifications and data are needed. Nevertheless, the possibility that the dragging spin force of an AGN BH jet is forcing and driving to a collinear jet blazing event might be a tool  for the discover of a very rare GW--$\gamma$ correlated signal. The present spin-spin correlation scenario, however, may occur mostly between AGN and medium-light BH, because of the very long $r_{s_{\mathrm{AGN}}}/c$ time of collapse, a long time needed to allow the spin-spin dragging forces to bend and align the small BH jet with the great one. This AGN time scale GWs are not observable with LIGO-VIRGO time scale like detector windows. We must in fact remind that the GWs  efficiency of any asymmetric mass system is very poor. Thus, there are no more good news (yet) regarding terrestrial GWs detectors, with the actual design such LIGO-VIRGO, inherent to events like 3C~279 one. We have to wait for future detectors such as LISA, located at terrestrial Lagrangian points, in order to reach for longer and longer GWs' time scale and corresponding larger masses BH-BH systems.
For those larger systems it may be the drag the one being able  to force the twin jet of the binary BHs toward Earth. Unfortunately, once again, the probability to observe  a nearly isotropic GW (which is quite high alone) together with the probability that such a system is simultaneously pointing at us, at the moment of the final collapse, is extremely poor for this kind of BH-BH ($\sim10^5$ M$_\odot$) systems. Only (or mostly) further detailed X-$\gamma$ minutes detections of  AGNs with short flares are the best tool to disentangle the processes and to confirm the eventual evidence of our spin-spin twin blazing model.

\section{Acknowledgments}

The authors gratefully acknowledge to Prof.s Amos Ori, Noam Soker, Robert Jantzen, Donato Bini for the very useful
discussions and suggestions. In particular the author D.F. is grateful to the Physics Depart. of Technion for the invitation and the hospitality.

\end{document}